\documentclass[conference]{IEEEtran}
\IEEEoverridecommandlockouts

\usepackage{cite}
\usepackage{amsmath,amssymb,amsfonts}
\usepackage{algorithmic}
\usepackage{graphicx}
\usepackage{textcomp}
\usepackage{subfigure}
\usepackage{amsmath}
\usepackage{booktabs, multirow} 
\usepackage{soul}
\usepackage[table]{xcolor} 
\usepackage{changepage,threeparttable} 
\usepackage{float}
\usepackage{hyperref}

\usepackage{soul} 

\newcommand{\randomsearch}{\textit{random-search}}
\newcommand{\gridsearch}{\textit{grid-search}}
\newcommand{\baeysianopt}{\textit{Bayesian optimization}}
\newcommand{\baeysianopts}{\textit{Bayesian optimizations}}

\def\BibTeX{{\rm B\kern-.05em{\sc i\kern-.025em b}\kern-.08em
    T\kern-.1667em\lower.7ex\hbox{E}\kern-.125emX}}
\begin{document}
\title{{Fast and Low-cost Search for Efficient Cloud Configurations for HPC Workloads}\thanks{We would like to thank CNPq (313012/2017-2), FAPESP(2013/08293-7), and Petrobras for their financial support to the realization of this research. Also, we would like to thanks the LMCAD for the provided infrastructure.
}
}

\author{\IEEEauthorblockN{Vanderson Martins do Rosario}
\IEEEauthorblockA{\textit{Institute of Computing} \\
\textit{UNICAMP}\\
Campinas, Brazil \\
vrosario@lmcad.ic.unicamp.br }
\and
\IEEEauthorblockN{Thais A. Silva Camacho}
\IEEEauthorblockA{\textit{Institute of Computing} \\
\textit{UNICAMP}\\
Campinas, Brazil \\
tcamacho@lmcad.ic.unicamp.br}
\and
\IEEEauthorblockN{Otávio O. Napoli}
\IEEEauthorblockA{\textit{Institute of Computing} \\
\textit{UNICAMP}\\
Campinas, Brazil \\
onapoli@lmcad.ic.unicamp.br}
\and
\IEEEauthorblockN{Edson Borin}
\IEEEauthorblockA{\textit{Institute of Computing} \\
\textit{UNICAMP}\\
Campinas, Brazil \\
borin@unicamp.br}
}

\maketitle

\begin{abstract}
The use of cloud computational resources has become increasingly important for companies and researchers to access on-demand and at any moment high-performance resources. However, given the wide variety of virtual machine types, network configurations, number of instances, among others, finding the best configuration that reduces costs and resource waste while achieving acceptable performance is a hard task even for specialists. Thus, many approaches to find these good or optimal configurations for a given program have been proposed. Observing the performance of an application in some configuration takes time and money. Therefore, most of the approaches aim not only to find good solutions but also to reduce the search cost. One approach is the use of Bayesian Optimization to observe the least amount possible of configurations, reducing the search cost while still finding good solutions. Another approach is the use of a technique named Paramount Iteration to make performance assumptions of HPC workloads without entirely executing them (early-stopping), reducing the cost of making one observation, and making it feasible to grid search solutions. In this work, we show that both techniques can be used together to do fewer and low-cost observations. We show that such an approach can recommend Pareto-optimal solutions that are on average 1.68x better than Random Searching and with a 6-time cheaper search.
\end{abstract}


\section{Introduction}

With the increasing access to computing resources using the cloud in the last years, many companies and researchers have migrated their computations from local infrastructure to the cloud. 
More than that, the ease of accessing powerful resources on-demand and at any time has empowered many academics that did not have easy access to accelerators, clusters, and other HPC infrastructures.

On the cloud computing model, users pay only for what they use; hence, they want to wisely use cloud resources, reducing their costs without compromising performance. For HPC workloads, the difference between choosing one computing resource over another can mean more than a 20-fold difference in total computation cost. Not surprisingly, solving the problem of selecting the most efficient cloud configuration for any given application became a relevant research topic in the last years.

Most of the first approaches to this problem consisted in creating or training performance models that describe the applications and the cloud resource\footnote{Most works focused on Virtual Machines resources.} performance. 
However, these techniques showed to be poor as they did not consider the dynamic performance of the cloud that is affected by the concurrent use and allocation of the physical resources, such as Virtual Machines or VMs. 
One possible solution is to dynamically try all possibilities and choose the best configuration; however, this approach can be expensive and hard to pay off. 
To reduce the cost of this search, tools such CherryPick~\cite{alipourfard2017cherrypick} and Arrow~\cite{hsu2018arrow} treat this problem as a black-box function optimization problem and employ \baeysianopt\ (BO) strategies to reduce the number of observations to find an efficient solution. 
This reduction significantly affects the search cost when looking for an efficient configuration and increases the set of applications that pay off such techniques.

Orthogonal to the approach of reducing the number of observations, Brunetta and Borin~\cite{jeferson} proposed the use of Paramount Iterations (PI)  to reduce the cost of each try itself. 
PI early-stops the execution of HPC programs but still collecting sufficient information to estimate the relative performance of different cloud configurations. 
However, so far, this approach has only been used together with \gridsearch.

In this work, we propose an approach that combines BO and PI to reduce the cost of searching for efficient cloud configurations even further.
We evaluate our approach searching for cost-efficient cloud configurations for 15 HPC workloads (5 kernels and 3 different input sizes) and showed a 6x reduction in average search cost when compared with not using PI.
The main contributions of this work are:
\begin{itemize}

    \item We proposed a novel approach to reduce the cost of searching for  efficient cloud configurations for HPC workloads by combining \baeysianopts\ (BO) and \emph{Paramount Iterations} (PI) together;

    \item We evaluate our approach with different BO techniques and show that the best technique may depend on the workload, defined by the application and its input data set.
    Nonetheless, the results indicate that the BO techniques perform consistently better than random and grid search.

    \item We discuss how to organize the search-space to reduce the number of dimensions and improve the performance of BO techniques.
    
    \item We argue that it may be essential to consider both the cost and performance of the configurations and propose a way of recommending a list of configurations that offer the best cost vs. performance trade-offs.
    
\end{itemize}

The remaining of the text is organized as follows: 
Section~\ref{sec:rw} discusses what has been done so far in this problem and maps our approach position in the literature; 
Section~\ref{sec:cloud_cfg_search} defines the problem and the BO details, mainly focusing in reducing the number of observations in the space; 
Section~\ref{sec:early-stopping} further describes PI and shows how it can be used to improve BO search cost even further; 
Sections~\ref{sec:exp_setup} and \ref{sec:exp_results} presents the setup of our experiments and the experimental results; 
Finally, Section~\ref{sec:conclusions} list our conclusions and future works. 

\section{Related Work}
\label{sec:rw}

Many public cloud providers exist such as Google Cloud, Amazon AWS, and Microsoft Azure, each one delivering tons of possible Virtual Machines (VM) and cluster configurations to be instantiated. For each of these instantiations and providers, there are different cost and performance associates. Prior work reports that it does not exist a one-size-fits-all VM type that is best for all workloads \cite{yadwadkar2017selecting, alipourfard2017cherrypick,no-one-fits, hsu2018arrow}. Thus, finding and matching the best cloud provider and best VM configuration to cost-efficiently run a program became an import problem that has been approached by many authors \cite{ernest,alipourfard2017cherrypick, yadwadkar2017selecting, hsu2018micky, hsu2018scout, hsu2018arrow, wu2019paraopt, jeferson, no-one-fits}. 

Some approaches tried to model cloud and program performance by extracting profiling information of both with multiple runs \cite{yadwadkar2017selecting, bodik2009automatic, ernest, 6655735, kunde2015workload, 8920852, 6676707}. The collected information is then used to optimally solve the matching problem offline or to train Machine Learning (ML) algorithms to learn how to recommend good cloud configurations. 
However, cloud performance is dynamic \cite{li2010cloudcmp} and can have a high variation in performance \cite{10.1145/2168836.2168847}, mostly because of the concurrent use of the physical resources from different VMs and users. Thus, the profiling collected from one run can become useless for a later run. 

Having that in mind, some authors start to dynamically explore the search-space of instance's configurations. Techniques such as \randomsearch\ or \gridsearch\ that extensively search the space of configurations could be used, but some works proposed the use of statistical methods such as \baeysianopt\ using Random Forrest (RF) \cite{hsu2018arrow} or Gaussian Process (GP) \cite{alipourfard2017cherrypick, hsu2018scout, wu2019paraopt} to decrease the number of configurations to be dynamically observed (an observation is defined as to run the program in a cloud configuration while measuring its execution time and cost). This approach reduces the search cost because it observes fewer configurations.

Although those techniques can reduce the number of instances to be observed they may still need a significant amount of resources (normally dollars) to run. Thus, most authors so far focus on big-data applications that are expected to continuously run for months and would supposedly repay the search cost \cite{alipourfard2017cherrypick,hsu2018arrow}. Hsu et al. \cite{hsu2018arrow} proposed the use of more low-level information collected while applying the dynamic Bayesian Optimization to augment the search and reduce the number of test instances, therefore, reducing the search-cost. In another approach, Hsu et al \cite{hsu2018micky}, achieve an 8.6-fold search-cost reduction by applying a collective-search and observing multiple applications at once.

On the other hand, orthogonal to the approaches of reducing the number of observations to reduce the search-cost, Brunetta and Borin \cite{jeferson} proposed an approach to reduce the cost of each observation itself. They run a \gridsearch\ (test all possible configurations), but using a technique called paramount iteration to extract the performance of HPC workloads with only a very small portion of their total execution. This early-stop technique showed to significantly reduce the number of resources need to find the best configuration.

No work so far, as far as we know, proposed the use of early-stop together with BO to minimize the search cost even further.

\section{Cloud Configuration Search}
\label{sec:cloud_cfg_search}

\subsection{Cloud Configuration Search Problem:}
Cloud computing has become an import tool for the scientific community, it allows the access of powerful computational resources without having to buy it or to manage it locally. This is specially important when the resources are only needed sporadically. Even institutes that promote scientific research and the cloud providers themselves are providing grants for running scientific experiments and simulation using the cloud. This has an even more special appeal for HPC workloads. Not all research institutes or researchers have easy access to HPC infrastructure or an HPC specialist to run their experiments.
    
With this migration from the use of local resources to cloud resources by researchers, new challenges arise. Questions such as, which cloud provider to use? What machines to hire? How to maximize performance or minimize the costs of the execution? All these become important and non-trivial to answer and several research work addressed them in the past few years. In this paper, we are interested in the problem of helping the cloud user to select the best cloud configuration to run HPC workloads in the Amazon Cloud (AWS). Notice, however, these results could be extended to any other cloud provider. 

The problem of searching for cloud configuration, minimizing a cost to run the experiments, could be described as 

\begin{equation}
    \min cost(cfg)
    \label{eq:mincost}
\end{equation}
where $cfg$ is a cloud configuration and cost is the amount of some resource used to run a program $p$ in that configuration. In the specific case of this paper $cfg$ can be defined as such:

\begin{equation}
    cfg = (vm, n) \in VM\times\{1,2,4,8,16,32\}
\end{equation}

where $vm \in$ VM, VM is the set of all possible VM configurations in a specific cloud provider and $n$ is the number of $vm$ instances used to form the cluster. The ordered pair $(vm, n)$ is referred in this paper as cloud configuration. So, for a given HPC workload $p$, we want to find the cloud configuration that minimizes the cost of executing it. 

This cost is calculated from the time ($T(p)$) to execute a program $p$, the price ($price(vm)$) of each $vm$, $n$, and a random noise $\epsilon$ that comes from cloud provider context and the measurement mechanism

\begin{equation}
    p => cost(cfg) = T(p) \times price(vm) \times n + \epsilon
    \label{eq:cost-from-t-and-price}
\end{equation}

Furthermore, we show that the list of Pareto-optimal configurations found during the search could form a list of recommendations of cloud configurations that dominates all others for both $T(p)$ and $cost(vm, n)$. This is usefull as the user may want to find a cloud configuration that minimizes the cost, but they are also concerned with the execution time.  

Finally, the search itself has a cost: the accumulated cost of all observations. Given a set of observations $O \subseteq VM\times\{1,2,4,8,16,32\}$, the search cost is $\sum_{i=0}^{|O|}{cost(O_{i})}$. A good search algorithm, therefore, solves (\ref{eq:mincost}) with the lowest possible search cost. 


\subsection{Approaches to the Cloud Configuration Search Problem}

The optimization problem characterized before in Equation~1 can be classified as a Black-Box Function Optimization Problem (BBFOP). A BBFOP is such that it has an interior function that can only be understood through exterior observation and experimentation, being those observations expensive in some resource and the function is not possible to be derivative no matter the amount of observations. Two straightforward ways to solve this optimization problem are the \gridsearch\ (test sequentially all possible inputs) and the \randomsearch\ (randomly test possible inputs until some predefined stop criteria). A more sophisticated common way to approach this problem is to use a Sequential Model-Based Optimization (SMBO) that tends to do need fewer observations to find optimal or almost optimal solutions.

SMBO consists of a model for the function being optimized that is updated after every observation using a prior-posterior Bayesian's approach. The Bayesian approach provides a posterior distribution of the black-box function, and estimates the uncertainty that help decides where to observe next, to find a maximum (or minimum). Frequent models used are Gaussian Process (GP) and Random Forrest (RF). An increasingly popular direction has been to use Gaussian Process (GP) to make smoothness assumptions on the function.

The observations used to update the model in SMBO are cherry-pick using information contained in the prior model. There are different strategies to pick the next point: to select the point that maximizes the probability of improvement (MPI) \cite{kushner1964new}; to select the point that maximizes the expected improvement(EI) \cite{movckus1975bayesian}; or to select the point that have the upper confidence bound (UCB) \cite{srinivas2009gaussian} on the maximum function value. Each algorithm reduces the black-box function optimization problem to a series of optimization problems of known acquisition functions.

The way of analyzing the performance of a BO approach differs too: objectives functions to analysis GP include cumulative regret, where every evaluation results in a reward or cost and the average of all function evaluations is compared to the maximum value of the function; simple regret that takes into account only the best value found so far \cite{bubeck2009pure}; the performance under a fixed finite budget \cite{grunewalder2010regret}; or the uncertainty about the location of the function maximizer \cite{hennig2012entropy}. For our problem of finding cloud configurations, more than solving Equation \ref{eq:mincost} we also want to solve it using the least amount of resources possible. Each observation in the cloud configuration space has its cost in time and dollars. Authors have used BBFOP to reduce the number of tests thus reducing the search total cost. 
However, for applications that only run one time or runs for a small amount of time, the price of the search can still surpass the gains by finding a good configuration.

Successful applications of BO includes the tuning of hyperparameters for complex models and algorithms in computer vision, machine learning and robotics \cite{brochu2010tutorial, calandra2014experimental,krause2011contextual,lizotte2007automatic,10.5555/2999325.2999464,thornton2013auto}. 
Recently the BBFOP field had a significant increase in interest for the hyperparameter tuning problem in Machine Learning. Machine Learn practitioners often approach the hyperparameter space search using brute-force methods like \randomsearch\ and \gridsearch\ \cite{10.5555/2188385.2188395}. In an effort to develop more efficient search methods, the problem of hyperparameter optimization has recently been dominated by Bayesian optimization methods \cite{10.5555/2999325.2999464,10.5555/2986459.2986743,10.1007/978-3-642-25566-3_40} that focus on optimizing hyperparameter configuration selection. 
Existing empirical evidence suggests that these methods outperform a \randomsearch\ \cite{thornton2013auto,eggensperger2013towards,snoek2015scalable}. 
However, these methods tackle the fundamentally challenging problem of simultaneously fitting and optimizing a high-dimensional, non-convex function with unknown smoothness, and possibly noisy evaluations. For high-dimensional problems, standard Bayesian optimization methods perform similarly to \randomsearch\ \cite{wang2013bayesian}.

\begin{figure}[!htb]
  \centering
  \subfigure[\textit{Random-search}.]{\includegraphics[scale=0.28]{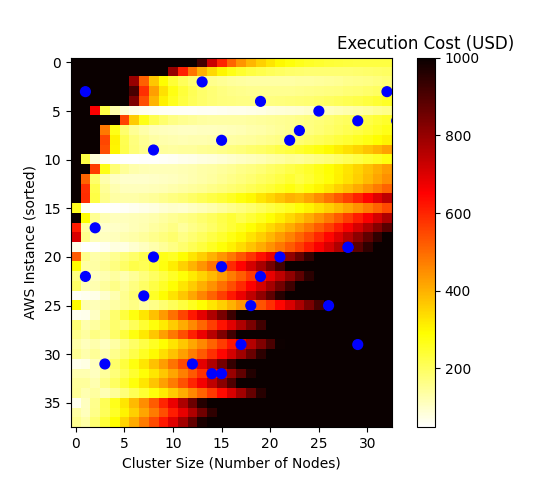}}
  \hspace{-1cm}
  \subfigure[Random Forest.]{\includegraphics[scale=0.28]{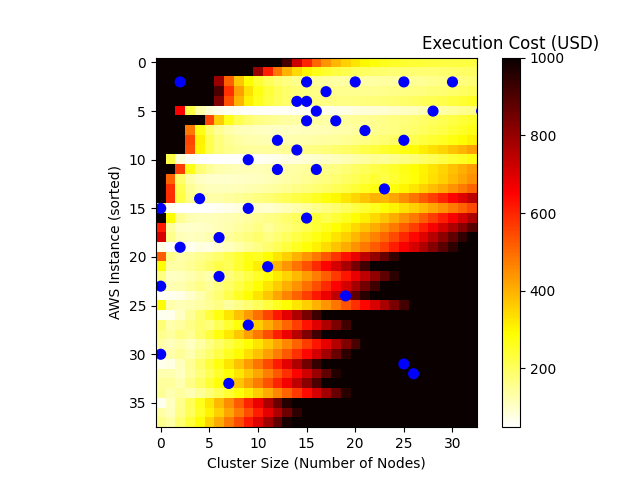}}
  \hspace{-0.5cm}
  \subfigure[GP-EI.]{\includegraphics[scale=0.28]{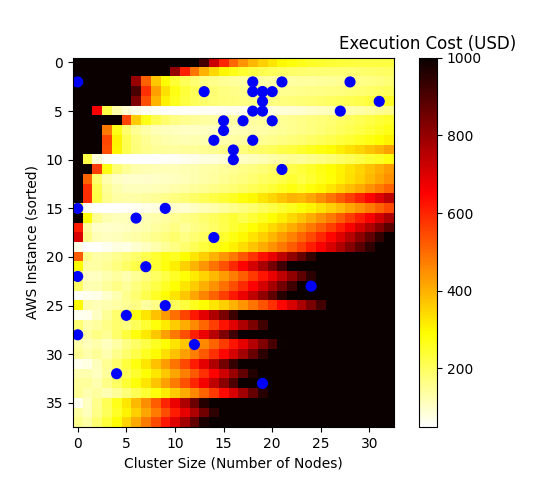}}
  \hspace{-0.5cm}
  \subfigure[GP-MPI.]{\includegraphics[scale=0.29]{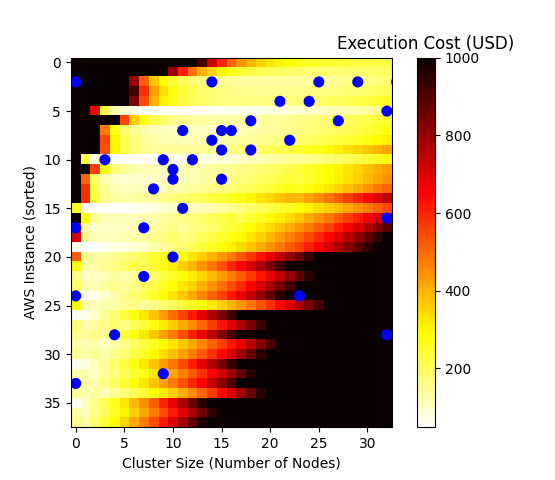}}
  \caption{Distribution of 32 observations in a model-generated space for 4 SMBO techniques.}
  \label{fig:searchexample}
\end{figure}

To exemplify SMBOs models and acquire functions results and how they observe different points in the space, we model a fictional cloud space search using AWS VMs information and price. We created an execution model inspired on Amdahl's Law and we ran 32 observations of \randomsearch\, Random-Forrest, GP-EI and GP-MPI. The result is shown in Figure~\ref{fig:searchexample}. Note that different from \randomsearch\, RF and GP tend to focus on good configurations (its able to learning through observations). Moreover, GP-MPI tends to more spatially explore the space while EI tends to maximize the best-found point by looking for nearby solutions. This is known as the ``exploitation vs exploration'' trade-off.

\section{Improving Search Cost by Early Stopping}
\label{sec:early-stopping}

Sophisticated techniques such as SMBO try to reduce the search's cost by reducing the number of observations on the search space.
In addiction to this approach, one may try to reduce the cost of the observations themselves. 

In the ML field, an orthogonal approach to hyperparameter optimization focuses on speeding up configuration evaluation. 
One of such approaches used in ML is early-stopping that can be used together with techniques such as Hyperband~\cite{li2017hyperband} to decide when to stop and how to allocate resources for each test. 

Brunetta and Borin~\cite{jeferson} showed that, in the context of HPC workloads on the cloud, it is possible to estimate the relative performance of different virtual machine types with very little execution time by only collecting information about few iterations of the main execution cycle of the application, or \emph{Paramount Iteration} (PI), as suggested by the authors.

In experiments, the authors found that collecting some, as small as one, of such iterations are enough to fully understand the final performance of the application. 
They used it to accelerate a \gridsearch\ of configurations in the Microsoft Azure Cloud. Executing only the first iteration of an HPC application and stopping it, reduces drastically the cost of the observations.

\begin{figure}[htb]
    \centering
    \includegraphics[scale=0.6]{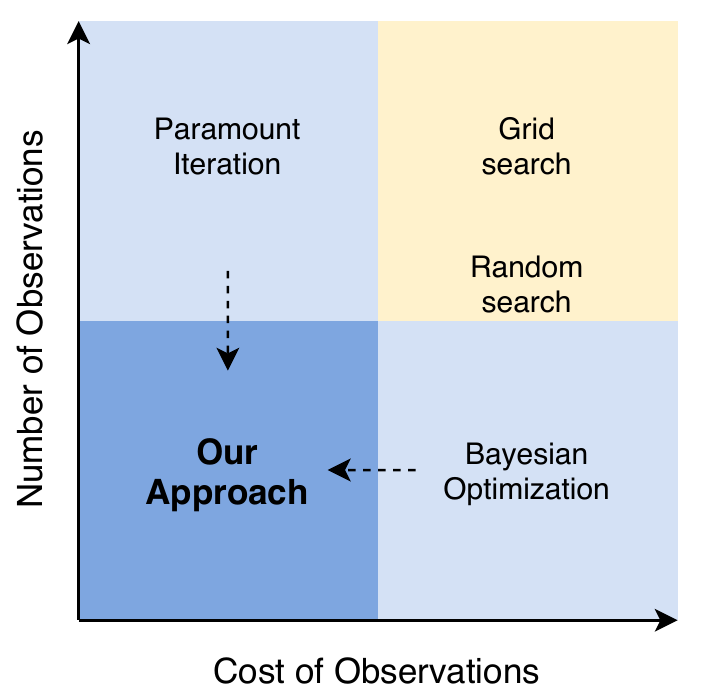}
    \caption{Diagram of cloud search approaches in a space of number of observations and cost of observations. Our approach combines the best of the two dimensions.}
    \vspace{-0.5cm}
    \label{fig:approaches}
\end{figure}

In this work, we present a coupling between the two approaches to reduce the search-cost (at least for HPC workloads). 
We reduced the number of observations instances using Bayesian Optimization and we use the \textit{Paramount Iteration} to perform an early-stop and make each observation faster and cheaper. 
Figure~\ref{fig:approaches} illustrates how the approaches used so far in the literature compare to ours.

\section{Experimental Setup}
\label{sec:exp_setup}

\subsection{Benchmark Suite}

To test our proposed approach, we execute the \textit{Numerical Aerodynamic Simulation Parallel Benchmarks} (\texttt{NPB})~\cite{Bailey2011} on AWS virtual machines. 
NPB is a suite of parallel computer performance benchmarks that has five kernels and three simulated computational fluid dynamics (CFD) applications. 
Each benchmark comes with a set of inputs that are divided into the following classes: \texttt{S}, \texttt{W}, \texttt{A}, \texttt{B}, \texttt{C}, \texttt{D}, \texttt{E}, and \texttt{F}. 
Class \texttt{S} consists of small workloads for quick tests; class \texttt{W} is designed to be executed on workstations; classes \texttt{A}, \texttt{B} and \texttt{C} contains standard workloads; and the classes \texttt{D}, \texttt{E} and \texttt{F} are composed of large workloads. 
In this work, we used \texttt{NPB} 3.4 with \texttt{MPI} support. 
We used the five kernels available with the input classes \texttt{C}, \texttt{D}, and \texttt{E}, a total of 15 workloads. 
Each of the used kernels are detailed in Table~\ref{tab:kernels}.

\begin{table}[!htb]
\centering
\tiny
\caption{Details of \texttt{NPB}'s kernels used in this work.}
\begin{tabular}{llll}
\hline
\multicolumn{1}{l}{\textbf{Benchmark}} & \multicolumn{1}{l}{\textbf{Description}}  & \multicolumn{1}{l}{\textbf{Language}} \\ \hline
EP & Embarrassingly Parallel & Fortran  \\
FT & Discrete 3D fast Fourier Transform & Fortran \\
MG & Block Tri-diagonal solver & Fortran \\
IS & Integer Sort, random memory access & C \\
CG & Conjugate Gradient & Fortran \\ \hline
\end{tabular}
\label{tab:kernels}
\end{table}

We instrumented the kernels' code to report the execution time of each Paramount Iteration (PI), and to stop the execution after executing 4 paramount iterations. 
The modified benchmarks can be found in our lab repository \footnote{Modified ES-NPB benchmarks: to be available after peer-review}. 

\subsection{Cloud Provider and Configurations' Space}

We chose the AWS cloud provider to execute our experiments. 
Nonetheless, it is worth mentioning that the techniques discussed here are not dependent on the cloud provider; hence, they can be easily used with other cloud providers.

AWS provides a wide variety of virtual machine (VM) types that one can instantiate. 
The list is organized in categories based on the characteristics of the physical machines that execute each VM type. 
Given the HPC context, we chose VM types from the compute-optimized (C5*), the general-purpose (m5*), and the memory-optimized (r5*) categories. 
Table~\ref{tab:aws} shows the characteristics of the 32 VM types selected for our experiments\footnote{Information about the currently available AWS VM types can be found on \url{https://docs.aws.amazon.com/AWSEC2/latest/UserGuide/instance-types.html}}. 

\begin{table}\centering
\caption{VM Parameters and Price.}\label{tab:aws}
\tiny
\begin{tabular}{p{0.7cm}p{0.15cm}p{0.15cm}p{0.27cm}p{0.27cm}}\toprule
\textbf{VM type} & \textbf{vCPU} &\textbf{MEM (GiB)} &\textbf{Network (Gbps)} &\textbf{Price (USD)} \\\midrule
m5n.large &2 &8 &25 &0.119 \\
m5n.xlarge  &4 &16 &25 &0.238 \\
m5n.2xlarge  &8 &32 &25 &0.476 \\
m5n.4xlarge &16 &64 &25 &0.952 \\
m5n.8xlarge &32 &128 &25 &1.904 \\
m5n.12xlarge &48 &192 &50 &2.856 \\
m5n.24xlarge &96 &384 &100 &5.712 \\
m5dn.large &2 &8 &25 &0.136 \\
m5dn.xlarge &4 &16 &25 &0.272 \\
m5dn.2xlarge &8 &32 &25 &0.544 \\
m5dn.24xlarge &96 &384 &100 &6.528 \\
c5.large  &2 &4 &10 &0.085 \\
c5.xlarge &4 &8 &10 &0.17 \\
c5.2xlarge  &8 &16 &10 &0.34 \\
c5.4xlarge  &16 &32 &10 &0.68 \\
c5.9xlarge  &36 &72 &10 &1.53 \\
\bottomrule
\end{tabular}
\quad
\begin{tabular}{p{0.7cm}p{0.15cm}p{0.15cm}p{0.27cm}p{0.27cm}}\toprule
\textbf{VM type} & \textbf{vCPU} &\textbf{MEM (GiB)} &\textbf{Network (Gbps)} &\textbf{Price (USD)} \\\midrule
c5.12xlarge  &48 &96 &12 &2.04 \\
c5.18xlarge  &72 &144 &25 &3.06 \\
c5.24xlarge  &96 &192 &25 &4.08 \\
c5n.large  &2 &5.25 &25 &0.108 \\
c5n.xlarge  &4 &10.5 &25 &0.216 \\
c5n.2xlarge  &8 &21 &25 &0.432 \\
c5n.4xlarge  &16 &42 &25 &0.864 \\
c5n.9xlarge  &36 &96 &50 &1.944 \\
c5n.18xlarge  &72 &192 &100 &3.888 \\
r5.large &2 &16 &10 &0.126 \\
r5.xlarge  &4 &32 &10 &0.252 \\
r5.2xlarge  &8 &64 &10 &0.504 \\
r5.4xlarge &16 &128 &10 &1.008 \\
r5.8xlarge  &32 &256 &10 &2.016 \\
r5.12xlarge  &48 &384 &10 &3.024 \\
r5.24xlarge  &96 &768 &25 &6.048 \\
\bottomrule
\end{tabular}
\end{table}

\vspace{0.2cm}

The cloud configurations search-space we explore in our experiments is composed of all pairs $(vm, n)$ where $vm$ indicates one of the 32 VM types in Table~\ref{tab:aws}, and $n\in\{1, 2, 4, 8, 16, 32\}$, indicates the number of instances used on the configuration, also referred as the cluster size. 
Hence, there are a total of 192 possible configurations per program. 
In real-life scenarios, this number can become much larger if we consider different providers and more VM type categories. 
Nonetheless, the search-space selected for our experiments is already challenging, since executing all NPB kernels to its completion on all configurations would cost more than 600~USD.

\subsubsection{Using Fewer Dimensions} other works~\cite{alipourfard2017cherrypick,hsu2018arrow} use VM's characteristics such as the number of vCPUs, memory size, among other, to create the dimensions of the search space.
However, as mentioned before, there is evidence that BO has a poor performance in high-dimensional spaces. 
Thus, we decide to use a 2-dimensional space to represent our problem. 
In contrast to using parameters of the VM types as dimensions, we use the VM type itself and the size of the cluster (i.e., the number of VM instances on the configuration). 

BO also tends to work better when the search-space is smooth. 
Therefore, to achieve that we sorted the VMs' dimension by their cost and amount of memory.
In most cases, this significantly improves the smoothness of the space by putting inexpensive machines that do not have the necessary amount of memory in one side and machines that are excessively expensive on the other side. 
With the most cost-effective solutions usually located between both, in a valley. 
In some configurations, for many reasons (lack of memory, vCPU number different than power of two, etc), some workloads does not work. When the search observe these configurations we consider the cost of the observations but we neither use the configuration in the solution or use it to update the Bayesian model. 

In spite of this has never been mentioned in the literature about cloud configuration search, we discovered from our experiments that without these transformations BO may perform similar to \randomsearch\ with low or no advantage over it. 

\subsection{Experimental Methodology}

Having our search-space and our benchmarks, we ran each of 5 NPB's kernels with each of the three input classes (C, D, and E) in each of the 192 cloud cluster configurations aforementioned. For each one of the 2880 executions (15 workloads times 192 configurations), we collect the execution time of the first four paramount iterations.

The kernels' performance on each of the 192 configurations is then used to compute the configurations cost using Equation~\ref{eq:cost-from-t-and-price}. 
Finally, once we have all cloud-cost-space calculated, we use the Python library GPyOpt~\cite{gpyopt2016} to simulate an online search in this search-space. 

As the searches are stochastic, we executed 50 times and collected their geometric mean.  
We tested the following  search approaches: \randomsearch\ and \baeysianopt\ (GP and RF) with MPI, EI, and LCB. 
We executed all searches with 32 observations (one per search iteration).
All \baeysianopt\ strategies have their first eight observations randomly picked to initiate the model.

The source code and all data collected from our experiments will be available in our lab repository\footnote{to be available after peer-review}.

\section{Experimental Results}
\label{sec:exp_results}

\subsection{Validation of Paramount Iteration for NPB Kernels}

\begin{figure}[!htb]
    \centering
    \includegraphics[scale=0.27]{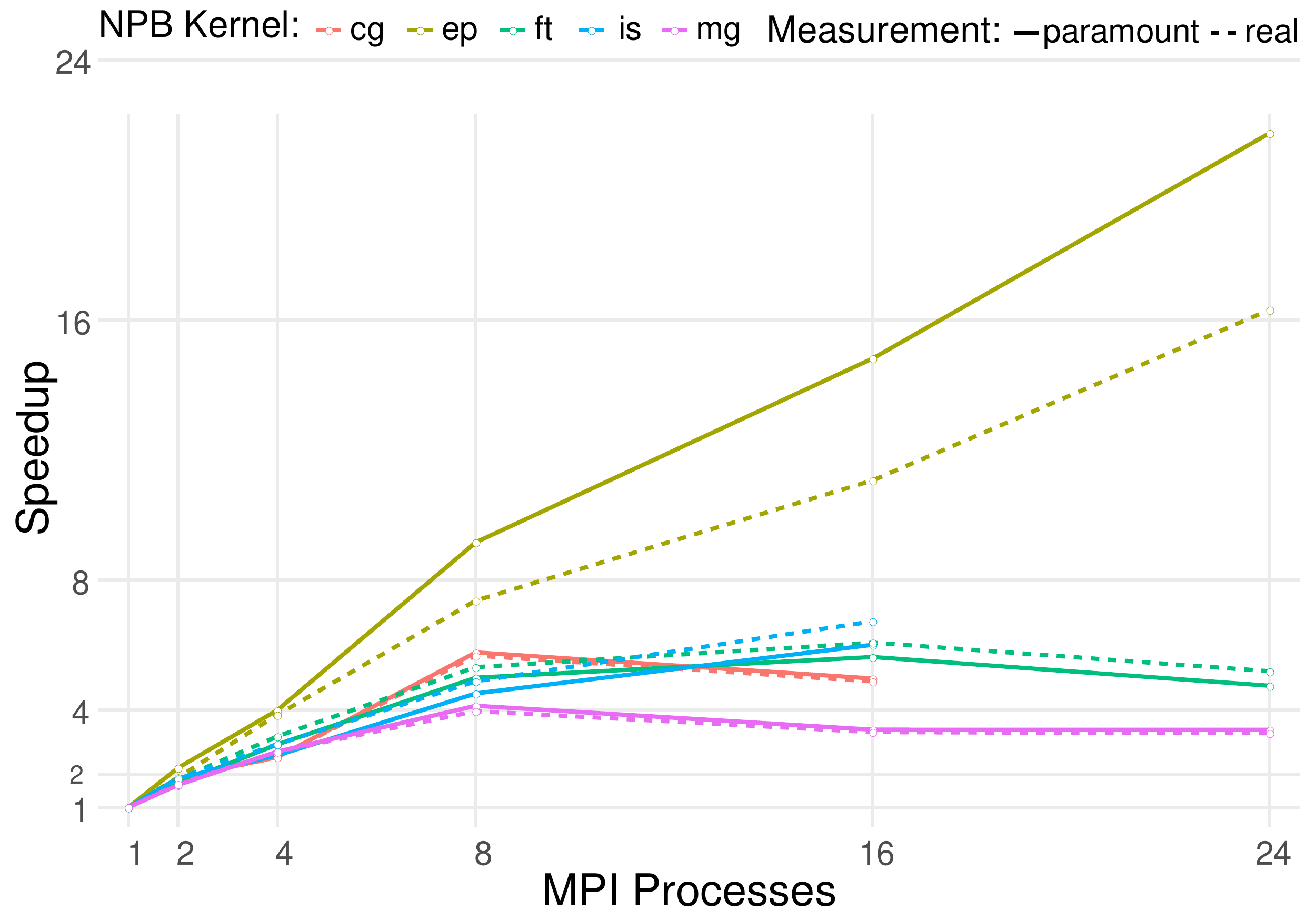}
    \caption{NPB Kernel's speedup measured with the average of 1 paramount iteration execution and with the real runtime for the whole execution. It shows less than 7.7\% difference between the two measurements. IS and CG only executes with power-of-two threads.}
    \label{fig:PIspeedup}
\end{figure}

Before executing the cloud experiments, we first used our local cluster to validate the Paramount Iteration (PI) concept, making sure that the first paramount iterations' performance could be used to estimate the performance of the whole execution on a given configuration. 
We used our modified version of the NPB with support to monitor the PIs to collect the first four iterations' execution time. 
We did this for every kernel using the input class E and varying the number o MPI process from 1 to 24 in a local machine with a Intel Xeon processor. 
We also collected the total execution time of these kernels using the original code, without the PI. 
Figure~\ref{fig:PIspeedup} shows the speedup foreseen by the PIs and the real speedup of the kernels.

Notice in the plot that the speedups estimated by the PIs follow the trend of the real ones. 
Moreover, the absolute value of the speedup is also close, less than 10\% distant from the average. EP result highlights as being the one with a larger distant between both measurements. We discover that this come from the fact that EP only have communication in the initialization and in the finalization, not in the PI. Thus, in the point of view of the PI measurement EP has an almost linear speedup, while in reality values have the communication overhead reducing its efficiency. However, despite that, even for EP both lines have the same behaviour, thus PI could still be used to compare performance for EP in difference configurations.

Thus, we have evidence that we can use the performance of the first PIs to compare the performance of different cloud configurations for the NPB kernels.  

\subsection{\textit{Grid-search} in AWS's configurations}

\begin{figure}[!htb]
    \centering
    \includegraphics[scale=0.42]{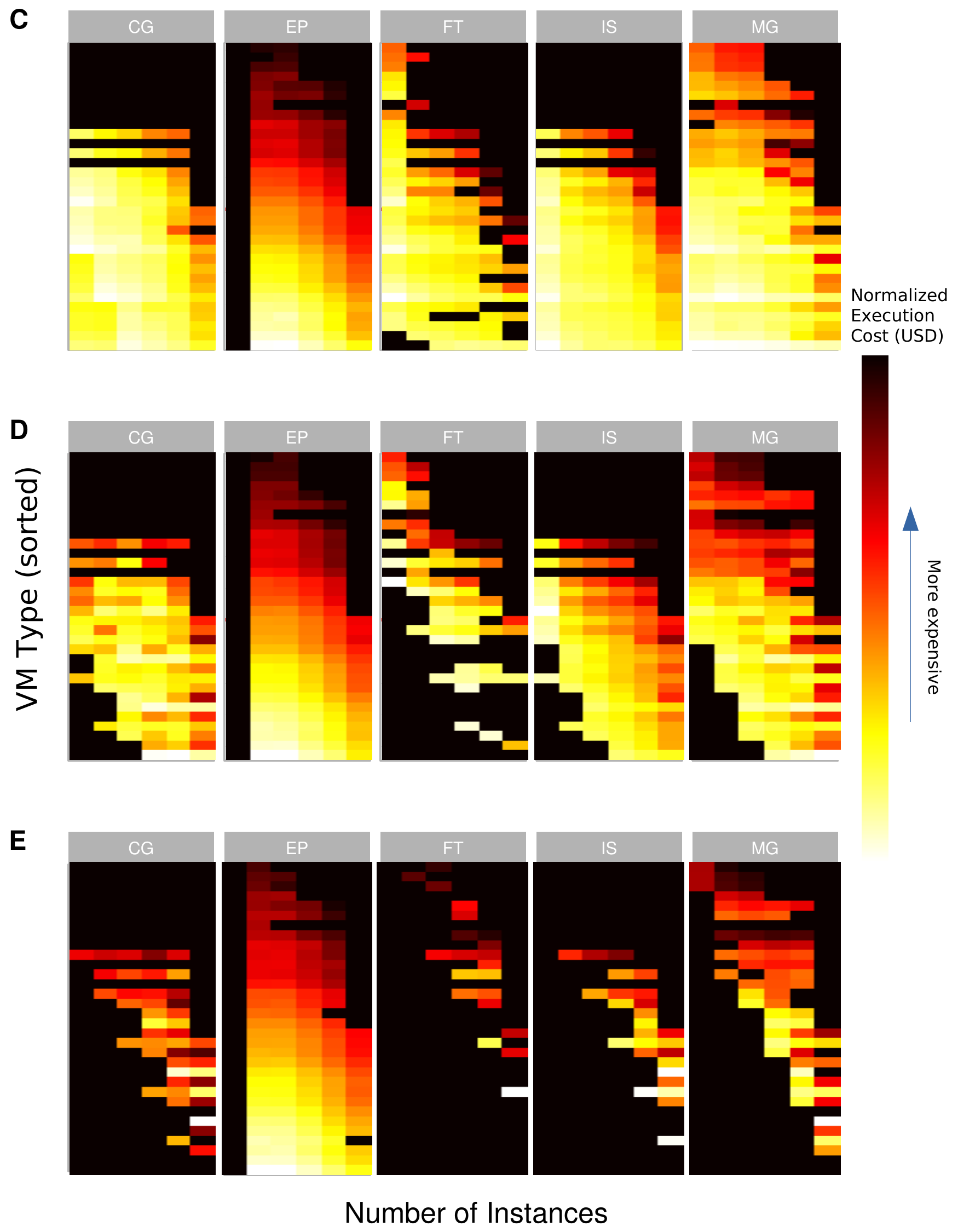}
    \caption{Real AWS cost space collected in our experiments for NPB kernels. The darker the color, the more expensive the configuration. AWS VM types and number of instances are sorted as in Figure~\ref{fig:searchexample}.}
    \label{fig:searchspace}
\end{figure}

As discussed in the previous section, we ran the NPB kernels with three input classes for each one of the selected cloud configurations to create our search-space. 
Figure~\ref{fig:searchspace} shows a heat map that indicates the cost of running each kernel with each input data set in each cloud configuration. 
It is possible to notice that when we increase the input data set (C$\rightarrow$E), increasing the demand for computation and memory, many configurations become unfeasible or too expensive (represented in black).
Thus, the larger the input, the fewer configurations provide good cost-benefit, and the higher the average cost of the configurations. 
This difference can also be noticed when we go from one benchmark to another (e.g., from MG-D to FT-D).
Therefore, both the application and its input data set have a significant impact on the search-space and, thus, implications on the search cost and results. 

\subsection{SMBO Techniques Comparison: no one rules them all}

\begin{figure}[!htb]
    \centering
    \includegraphics[scale=0.35]{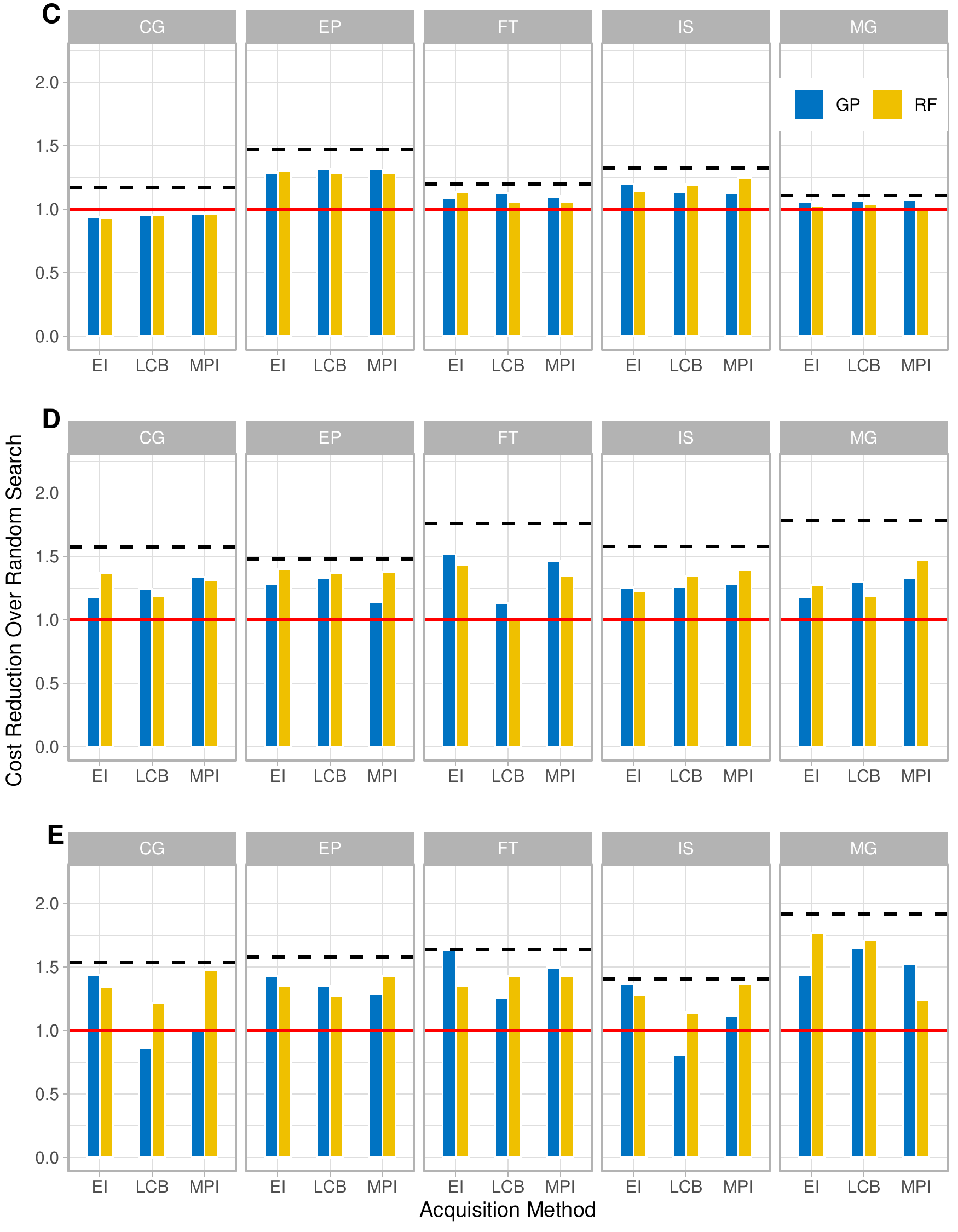}
    \caption{Average cost reduction achieved on 50 executions of each SMBO technique in the space from Figure~\ref{fig:searchspace}. Higher is better. The cost is normalized by the cost achieved by the Random Search approach, which is depicted by a red-straight line. Black-dashed lines shows the cost of the best possible solution.}
    \label{fig:performance}
\end{figure}

We applied the SMBO techniques to search the space depicted in Figure~\ref{fig:searchspace} to try to find the less expensive configurations in each space.
The results in Figure~\ref{fig:performance} show that no SMBO technique dominates in our experiments.
However, although it is visible that \baeysianopt\ approaches (BO) achieve better results than \randomsearch, no SMBO model or acquisition function seems to be better for most cases. 
Moreover, the results indicate that the best technique may depend on the search-space and, therefore, the program and input data set being analyzed.
Nevertheless, when considering the average of all acquisition functions, RF has a slight advantage over GP. 
Also, apart from the CG/C case, RF does not perform worse than the Random Search. 
These conclusions about RF being better than GP as a model for this problem is also pointed out by Hsu et al.~\cite{hsu2018arrow}.

The Black dashed line in the plots from Figure~\ref{fig:performance} shows the best possible solution for each case. 
Notice that SMBO techniques can find solutions close to the space-best while only iterating 16\% of the search-space (32/192). 
In other words, all SMBO approaches tested find solutions that are, on average, less than 13\% worst than the best solution that could be possibly found.

\subsection{Search Cost}

\begin{figure}[!htb]
    \centering
    \includegraphics[scale=0.45]{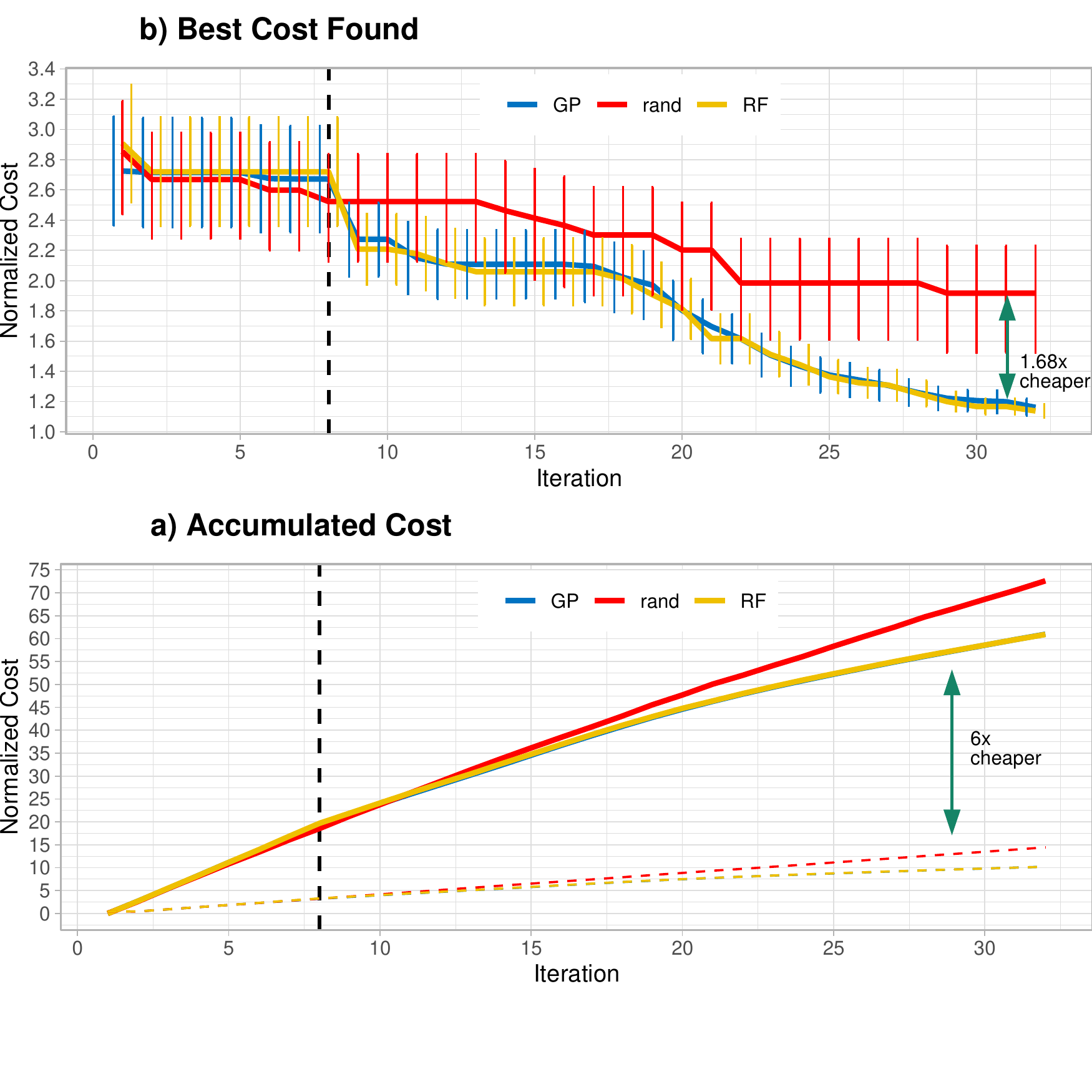}
    \vspace{-1cm}
    \caption{Best Cost and Accumulated Cost until each observation. Average values for all Kernels, Class and Acquisition. The dashed lines in Accumulated Cost represents the cost reduction when using PI.}
    \vspace{-0.3cm}
    \label{fig:convergence}
\end{figure}

Figure~\ref{fig:convergence}-a shows the normalized cost of the best configuration found by each search strategy after each observation.
We initiated the prior model space for all SMBO approaches with 8 random observations; hence, until the 8$^\text{th}$ observation, all SMBO approaches perform similar to the \randomsearch\ approach. 
However, it is possible to see a performance improvement just after the 9$^\text{th}$ observation between SMBO and \randomsearch.
After convergence, near the 15$^\text{th}$ observation, the best solution found by the SMBO techniques is 6x cheaper than when using \randomsearch. After the 20$^\textit{th}$ SMBO become significantly better than \randomsearch.

Figure~\ref{fig:convergence}-b shows the accumulated search cost after each observation for each search approach (the cost is normalized by the cheaper instance per kernel/class without PI).
Notice that, after the 8$^{th}$ observation, the accumulated cost for the SMBO approaches increases on a slower ratio. This happen because as SMBO lerns it tends to explore points in the space with more confidence to be cheaper, as illustrated on Figure~\ref{fig:searchexample}. Their search-cost were 36\% cheaper than \randomsearch\ after 32 observations.

When we apply the \textit{Paramount Iteration} (PI), we only need to execute four of the first paramount iterations of each application.
This means a significant reduction in execution time (and cost) when estimating the configuration performance.
On average, for the NAS kernels, we executed only 14\% of the total execution time of the applications to predict the relative performance of each configuration (4.43\% for CG, 17.33\% for FT, 12\% for MG, 40\% for IS and less than 1\% for EP). Larger the total execution time and the total amount of PIs, larger will be this difference between measuring only a first set of PIs and executing the whole program. In Figure~\ref{fig:convergence}-b we have the accumulated search cost after each observation with PI represented by dashed lines. 
As indicated in the figure, by combining PI with SMBO, we can reduce search-cost by a factor of 6-times. 


\subsection{Pareto-optimal Solutions}

As far as we are concerned, previous approaches to search the cloud configurations space focused only on one objective function. 
For example: reducing cost or improving performance.
However, in some cases, it is important to consider both objectives without combining them on a single objective function. 
In this context, it is useful to find the list of solutions that dominates all others in both objectives (Pareto-optimal solutions). 
Thus, the algorithm can recommend a list of solutions that vary between reducing the cost or the final runtime. 

\begin{figure}[!htb]
  \centering
  \subfigure[\textit{Random-search}.]
  {\includegraphics[scale=0.28]{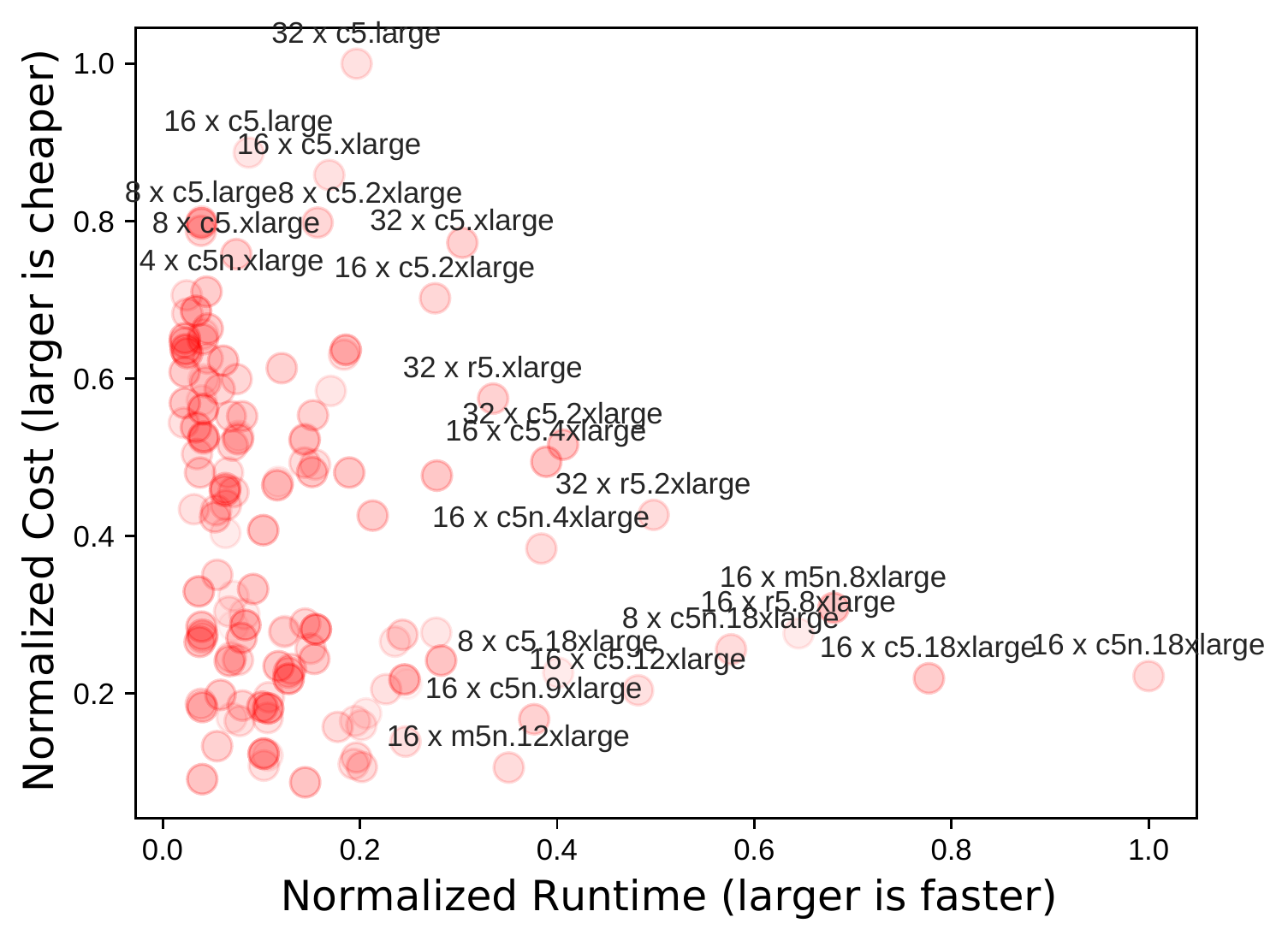}}
  \subfigure[GP-EI.]
  {\includegraphics[scale=0.28]{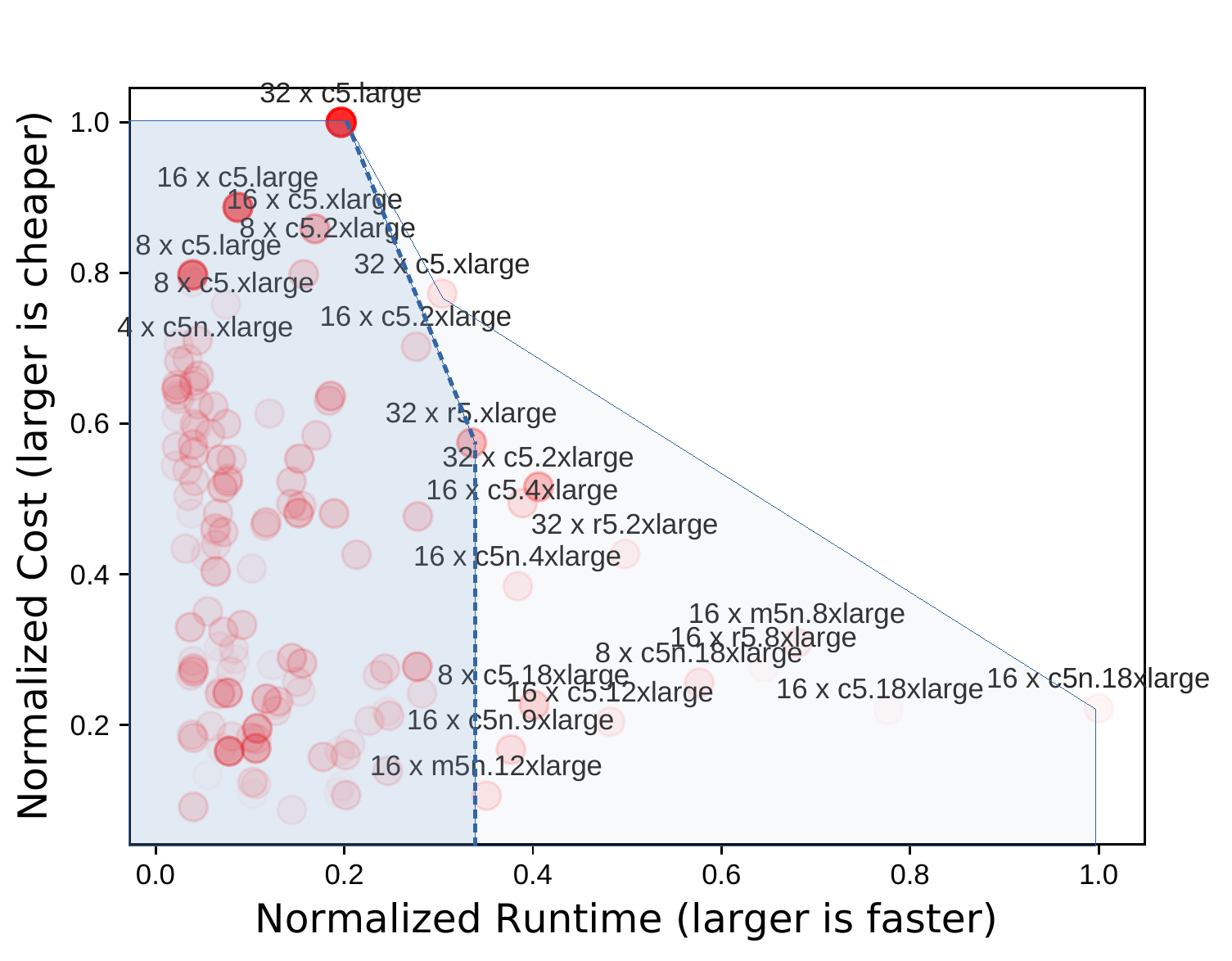}}
  \caption{Example of GP-EI and Random search result frequency with the Pareto Frontier delimited for GP-EI.}
  \label{fig:pareto}
\end{figure}

We scatter plot all possible configurations in a normalized runtime versus cost space where each point vary their opacity based on how many times they have being selected in 50 executions of the search. Figure~\ref{fig:pareto}-a does this for the \randomsearch\ and \ref{fig:pareto}-b for the GP-EI search. Both dimensions are normalized in such a wat that 1 represents the smallest cost or runtime and 0 is the largest. We can see from the plots that while \randomsearch\ tends to select all configurations, GP-EI tends to focus only on points that are highly cost efficient and ignoring points that are faster (on the extreme right). We plot the Pareto frontier for both the whole and the GP-EI most frequent configurations. Notice that following the points in the Pareto  we goes from cost-efficient to runtime-efficient configurations. However, as the plot shows, better objective functions could be use to make the search found better runtime-efficient points. What is a proposed future work. 

Two ways of measuring the performance of these recommendation lists are by the number of recommendations and their quality. 
The last can be measured by the area of the polygon formed by the Pareto-front \cite{1197687}.
We measure these for our experiments and RF showed to be slightly better than GP and LCB for both metrics.

\section{Conclusions}
\label{sec:conclusions}

In this work we propose a strategy that combines Sequential Model-Based Optimization (SMBO) techniques and the Paramount Iteration technique~\cite{jeferson} to search for efficient cloud configurations for HPC applications. 
We evaluate our approach using AWS computing resources and kernels from the NPB benchmark and conclude that SMBO approaches are far more efficient than Random Search in finding the good configurations after 32 observations (1.68x better results).
We also verified that different acquisition functions and models for SMBO have a small or no impact on its performance. 

Furthermore, we showed that the existing Paramount Iteration (PI) technique used to compare HPC workloads performance while early-stopping the application can be used to reduce search-cost without affecting its results. When using SMBO with PI, we achieved an 6-fold search-cost reduction. We also argue that it may be important to consider more than one solution from the search and propose a way of recommending a list of cost-efficient configurations to the final user. 
We then show two metrics to measure the recommendation quality and conclude that the \textit{Random Forrest} model has a slightly better result.

Finally, we point out that all proposed approaches in the paper can easily be used in real-world scenarios and adapted for other cost functions or cloud providers and used by companies and researchers to reduce their HPC experiments on the cloud.

\vspace{-0.5cm}
\bibliographystyle{IEEEtran}
\bibliography{IEEEabrv,mybibfile}

\end{document}